\documentclass[adp,fleqn]{w-art}
\usepackage[]{graphicx}
\chardef\bslash=`\\ % p. 424, TeXbook

\usepackage{wrapfig}
\newcommand{\bm}[1]{\mbox{\boldmath $#1$}}
\newcommand{\be}{\begin{equation}}
\newcommand{\ee}{\end{equation}}
\newcommand{\bea}{\begin{eqnarray}}
\newcommand{\eea}{\end{eqnarray}}
\newcommand{\la}{\langle}
\newcommand{\ra}{\rangle}
\newcommand{\di}{ {\rm d} }
\begin{document}

%%
%%  Most of the following commands will be completed by the publisher.
%%
%%  The copyrightyear is defined in the .clo file as the first argument
%%  of the copyrightinfo command. If the copyrightyear differs from that
%%  value it might be adjusted by the following definition:
%%
%% \renewcommand{\copyrightyear}{2002}% uncomment to change the copyrightyear.
%%
\DOIsuffix{theDOIsuffix}
%%
%% issueinfo for header and copyright line
%%
\Volume{12}
\Issue{1}
\Copyrightissue{01}
\Month{01}
\Year{2003}
%%
%%  First and last pagenumber of the article. If the option
%%  'autolastpage' is set (default) the second argument may be left empty.
\pagespan{1}{}
%%
%%    Dates will be filled in by the publisher. The 'reviseddate' and
%%    'dateposted' (Published online) entry may be left empty.
%%\Receiveddate{15 August 2004}
%%
%% \Reviseddate{30 November 1900}
%%
%%\Accepteddate{?? ???? 2004}
%\Dateposted{Published online 8 November 2004}
%%
%%
% up to three, separated by commas

%% \pretitle{Editor's Choice}

%% We have a short and a long form for the title. The short form
%% (optional argument) goes into the running head.

\title[Transversity]{Transversity in the chiral
quark--soliton model and \\
single spin asymmetries\footnote{The main results of this work
was obtained in collaboration with K.Goeke and P.Schweitzer,
Institut f\"ur Theoretische Physik II, Ruhr-Universit\"at Bochum,
Germany.}}

%% Please do not enter footnotes or \inst{}-notes into the optional
%% argument of the author command. The optional argument will go into
%% the header. If there is only one address the marker \inst{x} may be
%% omitted.

%%   Information for the first author.
\author[Efremov]{A.~V.~Efremov\footnote{
%Corresponding author \quad
E-mail: {\sf efremov@thsun1.jinr.ru}
     %, Phone: +49\,999\,999\,999, Fax: +49\,999\,999\,999}\inst{1}
     }}
\address[\inst{1}]{Joint Institute for Nuclear Research, Dubna 141980,
Russia}
%%
%%    Information for the second author
%\author[S. Author]{Second Author\footnote{Optional, e. g. e-mail}\inst{1,2}}
%\address[\inst{2}]{Second address}
%%
%%    Information for the third author
%\author[Th.\ Author]{Third Author\footnote{Optional, e. g. e-mail}\inst{2}}
%%
  \dedicatory{Dedicated to 60-th birthday of Prof. Klaus Goeke
in collaboration with whom most described results were obtained.}

\begin{abstract}
A short review of single spin asymmetries in deep inelastic
semi-inclusive processes, connected with prediction of chiral
quark-soliton model for the nucleon transversity distributions,
its possible theoretical understanding in the framework of
QCD-induced approach and arising difficulties is given.
\end{abstract}
\maketitle

\section{Introduction}
\label{sect1}

It is well known that three most important (twist-2) elements of
%PS
the
%PS
parton density matrix in a nucleon are the non-polarized parton
distributions functions (PDF) $f_1(x)$, the longitudinal spin
distribution  $g_1(x)$ and the transverse spin distribution
(transversity) $h_1(x)$ \cite{transversity}. The first two have
been successfully measured experimentally in
classical deep inelastic scattering (DIS) experiments but the
measurement of the last one is especially difficult since it
belongs to the class of the so-called chiral-odd structure
functions and can not be seen there.

The non-polarized PDF's
%PS was
have been measured for decades and are rather well
known in wide range of $x$ and $Q^2$.
%PS Its
Their behavior in $Q^2$ is well
described by the QCD evolution equation and serves as one of
the main
%PS source
sources of $\alpha_s(Q^2)$ determination.

The longitudinal spin PDF's
%PS draw
drew common attention during last
decade in connection with the famous "Spin Crisis", i.e.
astonishingly small portion of the proton spin carried by quarks
(see \cite{ael} and references therein).  The most popular
explanation of this phenomenon is large contribution of the gluon
spin $\Delta G(x)$.  The direct check of this hypothesis is one
of the main
%PS problem
problems of running dedicated experiments like COMPASS
at CERN and RHIC at BNL. Even now, however, there
%PS are
is some indication to a considerable value of $\Delta G(x)$ coming from
the $Q^2$ evolution of the polarized PDF's \cite{leader99} and
from the first direct experimental probe of $\Delta G(x)$ by
HERMES collaboration \cite{hermes00} with the result $\Delta
G(x)/G(x)=0.41\pm0.18$ in the region $0.07<x<0.28$. The latter is
in reasonable agreement with large $N_c$ limit prediction
\cite{efremov00} $\Delta G(x)/G(x)\approx 1/N_c$ for not very
small $x$.

Another problem here is the sea quark spin asymmetry. It is
usually assumed in fitting the experimental data that $\Delta\bar
u=\Delta\bar d =\Delta\bar s$. This {\it ad hoc} assumption
however contradicts with large $N_c$ limit prediction $\Delta\bar
u\approx-\Delta\bar d$ \cite{efremov00}. This was previously
discovered
%PS for
in the instanton model \cite{DK} and supported by
calculations in the chiral quark soliton model ($\chi$QSM)
\cite{Diakonov:1996sr,Dressler00}. An indication to a nonzero
value for $\Delta \bar{u} -\Delta \bar{d}$  was also observed in
\cite{MY}.

Concerning the transversity distribution it was completely
unknown experimentally till recent time. The only information
comes from the Soffer inequality \cite{Soffer:1995ww}
$|h_1(x)|\le{1\over2}[f_1(x)+ g_1(x)]$ which follows from density
matrix positivity. To access these chiral-odd structure functions
one needs either to scatter two polarized protons and to measure
the transversal spin correlation $A_{NN}$ in Drell-Yan process
that is the problem for running RHIC and future PAX (GSI)
experiments\footnote{ For predictions see \cite{Efremov:2004qs}
and references therein.} or to know the transverse polarization
of a quark scattered from transversely polarized target. There
are several ways to do this:
\begin{enumerate}
\item
To measure the polarization of a self-analyzing hadron to which
the quark fragments in a semi inclusive DIS (SIDIS), e.g.
$\Lambda$-hyperon.  The drawback of this method however is a
rather low rate of quark fragmentation into $\Lambda$-particle
($\approx 2\%$) and especially that it is mostly sensitive to
$s$-quark polarization. Also the polarization transfer from
parton to $\Lambda$-hyperon is unknown.
\item
To measure a transverse handedness in multi-particle parton
fragmentation \cite{hand}, i.e. the correlation of the parent quark spin
4-vector $s_\mu$ and jet particle momenta $k_i^\nu$,
$\epsilon_{\mu\nu\sigma\rho}s^\mu k_1^\nu k_2^\sigma k^\rho$
($k=k_1+k_2+k_3+\cdots$ is a jet 4-momentum).
\item
To use a new spin dependent T-odd parton fragmentation function
(PFF) \cite{Mulders:1995dh,Boer:1997nt,muldz} responsible for the
left-right asymmetry in one particle fragmentation of
transversely polarized quark relative to the quark momentum--spin
plane. (The so-called "Collins asymmetry" \cite{collins}.)
\end{enumerate}

The last two methods are comparatively new and only in the last
years some experimental indications to the transversal handedness
\cite{czjp99} and to the T-odd PFF \cite{todd} have appeared.

Concerning the new PDF's and PFF's. Analogous of PDF's $f_1,\
g_1$ and $h_1$ are the PFF's $D_1,\ G_1$ and $H_1$, which
describe the fragmentation of a non-polarized quark into a
non-polarized hadron  and a longitudinally or transversely
polarized quark into a longitudinally or transversely polarized
hadron, respectively. These PFF's are integrated over the
transverse momentum ${\bm p}_{h\perp}$ of a hadron with respect
to a quark. With ${\bm p}_{h\perp}$ taken into account, new PFF's
arise. Using the Lorentz- and P-invariance one can write in the
leading twist approximation 8 independent spin structures. Most
spectacularly it is seen in the helicity basis where one can
build 8 twist-2 combinations, linear in spin matrices of the
quark and hadron {$\bm\sigma$}, ${\bm S}$ with momenta ${\bm
k'}$, ${\bm p_h}$.

These PFF can be used to extract the information on the proton
transversity distribution from azimuthal asymmetries in SIDIS
with hadron production (pions and kaons) on a polarized nucleon
target
\be
\label{reaction}
l+\vec N \to l'+ h + X
\ee
recently observed by HERMES
\cite{Airapetian:1999tv}--\cite{Avetisyan:2004uz} and CLAS
\cite{Avakian:2002qp,Avakian:2003pk} collaborations.

In this short review I present the results of works
\cite{Efremov:2000za}--\cite{Efremov:2002ut} on the SIDIS
asymmetries that are connected with the transversity
distributions predicted in $\chi$QSM \cite{Schweitzer:2001sr}.

%%%%%%%%%%%%% Sect. kinematics & asymmetries %%%

\section{SIDIS kinematics and azimuthal asymmetries}
\label{sect-kin-asym}

In the framework of the parton model the squared matrix element
modulus of the  process (\ref{reaction}) is represented by
\begin{figure}[htb]
\begin{minipage}[t]{.40\textwidth}
\includegraphics[width=0.85\textwidth]{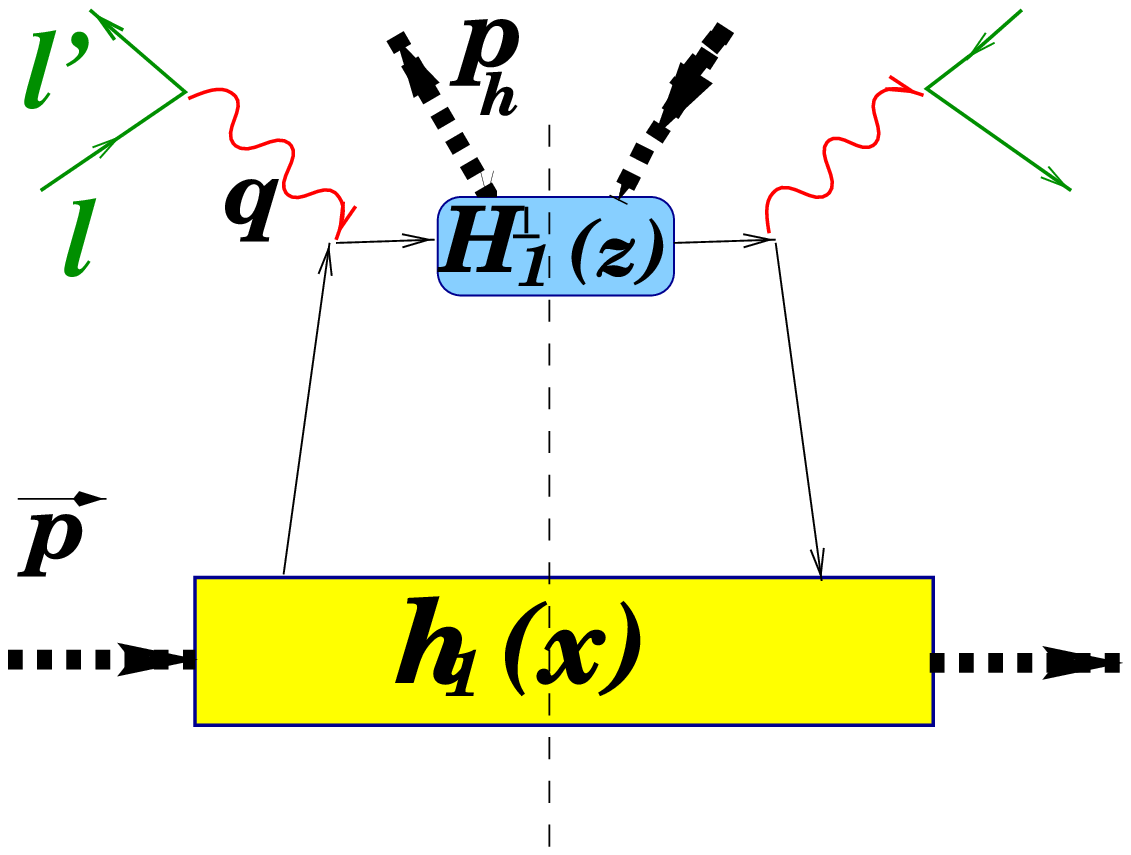}\\[-10mm]
\caption{The squared  modulus of the matrix element of the
process (\ref{reaction}) in the parton model summed over states
$X$. The $H_1^\perp(z)$ and $h_1(x)$ are  examples of PFF and
PDF, respectively.}
\label{diagram}
\end{minipage}
\hfil
\begin{minipage}[t]{.55\textwidth}
\raisebox{-7ex}{\includegraphics[width=
\textwidth,height=45mm]{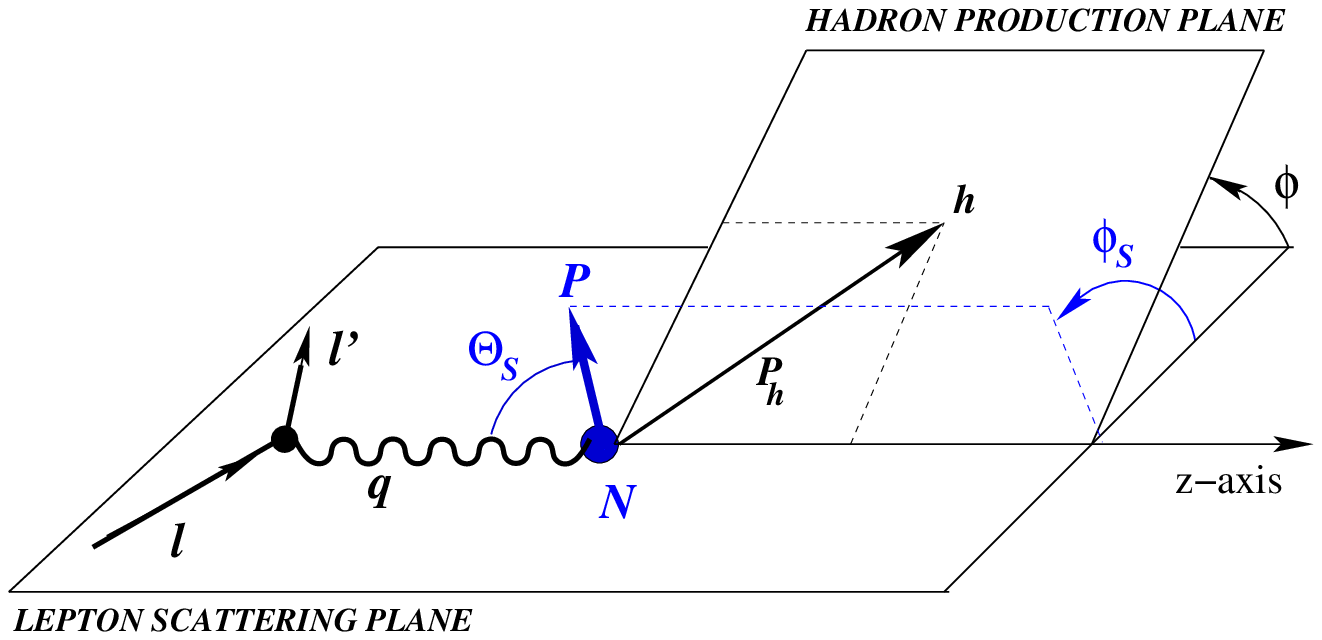}} \caption{Kinematics of
the process (\ref{reaction}).} \label{kinsidis}
\end{minipage}
\end{figure}
the diagram in Fig.\ref{diagram}  and can be written as a sum of
products of $x$-dependent quark distribution functions in a
nucleon, $x=\frac{Q^2}{2p\cdot q}$, with $q=l-l'$, $Q^2=-q^2$,
and $z$-dependent quark fragmentation functions of scattered
quark into hadron $h$, $z=\frac{p\cdot p_h}{p\cdot q}$. The
kinematics of the process (\ref{reaction}) is presented in
Fig.\ref{kinsidis}.

The cross section of the semi-inclusive production  of hadrons by
polarized leptons on the polarized target is a linear function of
the longitudinal lepton beam polarization, $P_l$, and the
target polarization, $P$, with  longitudinal,  $P_L$, and
transversal, $P_T$, components relative to a virtual photon
momentum $\vec q$ in laboratory r.f.:

\be
\label{xsect}
d\sigma=d\sigma_{00}+P_l d\sigma_{L0}+
P_L(d\sigma_{0L}+P_l d\sigma_{LL})
+|P_T|(d\sigma_{0T}+P_l d\sigma_{LT}).
\ee

For the target polarization $P_\|$, longitudinal relative to
the lepton beam, the transverse component is equal to
$|P_T|=P_\|\sin\theta_\gamma$, where $\theta_\gamma$ is the angle
of the virtual photon momentum $\vec q$ relative to the
lepton beam,
\be
\label{PT}
\sin\theta_\gamma\approx
2\frac{M}{Q}x\sqrt{1-y}\,,
\ee
where $y=\frac{p\cdot q}{p\cdot l}$ and $M$ is the nucleon mass.

In the parton model each of the partial cross
%PS section
sections contributing to the Eq. (\ref{xsect}) is characterized by the
specific dependence on the azimuthal angle of an outgoing hadron,
$\phi$, and on the azimuthal angle of transversal component of
the target polarization vector\footnote{For longitudinal target
polarization $P_\|$ the angle $\phi_S=0$ or $\pm\pi$}, $\phi_S$,
relative to the lepton scattering plane (see
Fig.\ref{kinsidis}) times the definite product of PDF and PFF
summed over quark and antiquark flavor times its charge squared.
Namely, contributions to (\ref{xsect}) for each quark and
antiquark flavor up to the order ${\cal O}(M/Q)$) have the
forms\footnote{We use the notations of the
work \cite{Mulders:1995dh,Boer:1997nt,muldz}. The letters
$g_1(G_1),\ h_1(H_1),\ f_1(D_1)$ indicate twist-2 PDF (PFF) with
longitudinally, transversally polarized or unpolarized partons,
%PS subscribe
subscripts $L,\ T$ indicate the polarization of hadron and
%PS superscribe
superscripts $\perp$ indicates
%PS to
a $p_\perp$-dependence. Note that very recently further structures
have been introduced and discussed \cite{Afanasev:2003ze} that could
contribute the longitudinally polarized cross sections. }
\cite{Mulders:1995dh,Boer:1997nt}:

\vspace{10mm}
\be
\label{pdfpff}
~
\ee

\vspace{-20mm}
\begin{tabular}{ll}
$d\sigma_{00}\propto$& $ xf_1(x)D_1(z)+
      xh_1^\perp(x) H_1^\perp(z)\cos2\phi -
      \frac{M}{Q}x^2f^\perp(x) D_1(z)\cos\phi$\,, \\
$d\sigma_{L0}\propto$& $\frac{M}{Q}x^2e(x)H_1^\perp(z)\sin\phi
+\frac{M}{Q}x^2h_1^\perp(x)E(z)\sin\phi$\,, \\
$d\sigma_{0L}\propto$&
{$xh_{1L}^\perp(x)
    H_1^\perp(z)\sin 2\phi $}
    {$+\frac{M}{Q}x^2h_{L}(x)
    H_1^\perp(z)\sin\phi$}\,,\\
$d\sigma_{LL}\propto$& $xg_1(x)D_1(z)+
     \frac{M}{Q}x^2g^\perp_L(x)D_1(z)\cos\phi$\,, \\
$d\sigma_{0T}\propto$&
{$xh_1(x)H_1^\perp(z)\sin(\phi+\phi_S)$}
       $+xh_{1T}^\perp(x) H_1^\perp(z)\sin(3\phi-\phi_S)$\\
    &{$+xf_{1T}^\perp(x)
      D_1(z)\sin(\phi-\phi_S)$}\,,\\
$d\sigma_{LT}\propto$& $xg_{1T}(x)D_1(z)\cos(\phi-\phi_S)$\,,\\[5mm]
\end{tabular}

\noindent
where
\begin{itemize}

\item[]
$f_1(x)\equiv q(x)$ is PDF of non-polarized quarks in a
non-polarized target,

\item[]
$g_1(x)\equiv\Delta q(x)$ is PDF of the longitudinally polarized
quarks in the longitudinally polarized target,
\item[]$g_{1T}(x)$
is the same as $g_1(x)$ but in the transversally polarized
target,

\item[]
$h_1(x)$ is PDF of the transversally polarized quark with
polarization  parallel to that one of a transversally polarized
target (so-called transversity),

\item[]
$h^{\perp}_{1L,T}(x)$ is PDF of the transversally polarized quark
with polarization perpendicular to the hadron polarization in the
longitudinally or transversally polarized target,

\item[]$h^{\perp}_1(x)$
is PDF of the transversally polarized quark in the non-polarized
target,

\item[]$f^{\perp}_{1T}(x)$
is PDF responsible for a left-right asymmetry in the distribution
of the non-polarized quarks in the transversally polarized target
(so-called  Sivers PDF \cite{Sivers:1989cc}),

\item[]$D_1(z)$
is PFF of the non-polarized quark in the non-polarized or
spinless produced hadron,

\item[]$H^{\perp}_1(z)$
is PFF responsible for a left-right asymmetry in the
fragmentation of a transversally polarized quark into a
non-polarized or spinless produced hadron (so-called Collins
PFF \cite{collins}).
\end{itemize}

The others ($E,\ e,\ g_L^\perp,\ h_L,\ f^\perp$) are the twist-3
functions entering with a factor $M/Q$. They have no definite
probabilistic interpretation, but are connected (except for
$e(x)$ and $E(z)$) to the above listed functions by the
approximate integral relations of the Wandzura-Wilczek type. For
example \cite{Mulders:1995dh,Boer:1997nt}

\be
\int d^2k_\perp\left(\frac{k_\perp^2}{2M^2}\right)h_{1L}^{\perp}(x,k_\perp)
\equiv h_{1L}^{\perp(1)}(x)=
-(x/2)h_L(x)=-x^2\int\limits_x^1d\xi h_1(\xi)/\xi^2.
\label{wwform}
\ee
Besides, in formula (\ref{pdfpff}) each term should be multiplied
by known kinematic factors depending on $y$, $\la k_\perp\ra$ and
$\la p_{h\perp}\ra$ (assuming Gaussian distribution) which are
omitted for simplicity.

The azimuthal asymmetries are defined as
\be\label{asim}
A_{BA}^{W(\phi,\phi_S)}(x,z,h) =
\frac{\displaystyle
\int\!\!\di y\,\di\phi\,\di\phi_S\,W(\phi,\phi_S)\,\left(
\frac{1}{P_+}\,\frac{\di^4\sigma_D^+}{\di x\,\di y\,\di z \di\phi}-
\frac{1}{P_-}\,\frac{\di^4\sigma_D^+}{\di x\,\di y\,\di z \di\phi}\right)}
{\;\;\;\;\;\;\;\displaystyle
\frac{1}{2}\int\!\!\di y\,\di\phi\,\left(
\frac{\di^4\sigma_D^+}{\di x\,\di y\,\di z\di\phi}+
\frac{\di^4\sigma_D^-}{\di x\,\di y\,\di z\di\phi}\right)}\;\;,
\ee
where $W(\phi,\phi_S)$ is an angular dependent weight from Eqs.
(\ref{pdfpff}) and $P_\pm$ denotes the target polarization
modulus. The subscripts B and A are 0, L or T for the
unpolarized, longitudinally or transversally polarized beam or
target (relative to the virtual photon direction).

It is clear that for the transversal target polarization one can
separate all the
%PS contained in (\ref{pdfpff}) components
components contained in (\ref{pdfpff})
carrying out the Fourier-analysis with respect to the angles $\phi$ and
$\phi_S$ plus change of the polarization sign (using
anti-symmetrization and symmetrization). For example, for
separation of the term with the pure transversity contribution
$h_1(x)H_1^\perp(z)$ it is enough to calculate the average value
$\la\sin(\phi+\phi_S)\ra$ and for separation of the  Sivers
function $f_{1T}^{\perp}(x)D_1(z)$ it is enough to find the
average value $\la\sin(\phi-\phi_S)\ra$.

For the longitudinal target polarization the total separation is
impossible since $\phi_S=\pm\pi$ or $0$ (see Fig.\ref{kinsidis})
and several different mechanisms can produce the
$\sin\phi$-asymmetry.  However it is possible to single out the
function $h_{1L}^\perp(x)$,  which is connected with transversity
via Eq. (\ref{wwform}), by measuring $\la\sin(2\phi)\ra$.

%%%%%%%%%%%%%%% Sect. Collins PFF %%%%%%
\section{Collins PFF}
\label{sect-Collins}

As it is seen from (\ref{pdfpff}) the Collins PFF that describes a
left--right asymmetry in the fragmentation of a transversely
polarized quark is especially interesting since it enters any
term connected with the transversity. The corresponding term of
fragmentation has the structure
$$
H_1^\perp\mbox{\bm\sigma}({\bm k'}\times
{\bm p}_{h\perp})/k'\la p_{h\perp}\ra,
$$
where $H_1^\perp$ is a function of the longitudinal momentum
fraction $z$ and hadron transverse momentum  $p_{h\perp}$. The
$\la p_{h\perp}\ra$ is the averaged transverse momentum of the
final hadron\footnote{Notice different normalization factor
%PS compare
compared with \cite{muldz}, $\la p_{h\perp}\ra$ instead of
$M_h$.}. Since the $H_1^\perp$ term is chiral-odd, it makes
possible to measure the proton transversity distribution $h_1$ in
semi-inclusive DIS from a transversely polarized target by
measuring the left-right asymmetry of forward produced pions. The
ratio $H_1^{\perp}\over D_1$ serves as analyzing power of
the Collins effect.

The problem is that, first, this function was completely unknown
till recent time both theoretically and experimentally. Second,
the function $H_1^\perp$ is the so-called T-odd fragmentation
function: under the naive time reversal ${\bm p_h},\ {\bm k'},\
{\bm S}$ and $\bm\sigma$ change sign, which demands a purely
imaginary (or zero) $H_1^\perp$ in the contradiction with naive
hermiticity. This, however, does not mean the break of
T-invariance but rather the presence of an interference of
different channels in forming the final state with different
phase shifts, like in the case of single spin asymmetry
phenomena\footnote{In this aspect they are very different from
the T-odd PDF's like $f_{1T}^\perp$ or $h_1^\perp$ which can not
exist since they are purely real. Interaction among initial
hadrons which could brings an imaginary part breaks the
factorization and the whole parton picture. Recently however it
was stated \cite{BMT} that effectively the necessary imaginary
phase shift can appear due to propagation of the scattered parton
in gluon field of the nucleon remnant. Since this phase shift
depends on the subprocess the corresponding PDF is, generally
speaking, not universal. In particular, it was shown that the
Sivers PDF in one loop approximation, where its factorization
seems proven \cite{Ji:2004wu,Collins:2004nx}, should have
opposite sign in Drell-Yan and SIDIS processes. Also this
functions are suppressed in $\chi$QSM (see footnote
\ref{chiral-t-odd}).} \cite{gasior}. A calculations of this
function in simple perturbative chiral theory can be found in
\cite{Bacchetta:2002es}.
\begin{wrapfigure}{R}{6.0cm}
\vspace{-4mm}
\includegraphics[width=6.2cm,height=6.0cm]{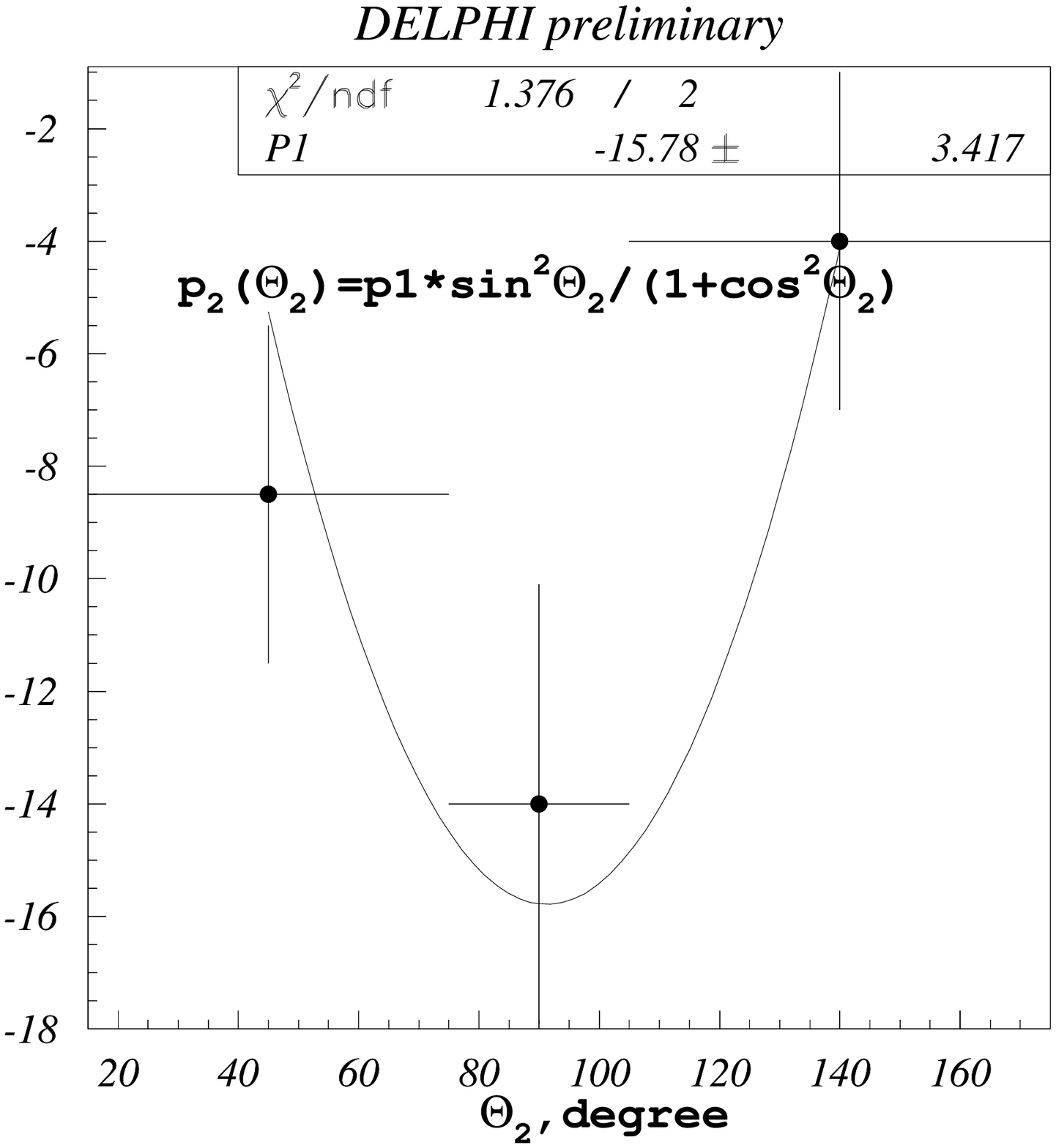}\\[-5mm]
\caption{ The $\theta_2$-dependence of the value
$P_2={6\over\pi}\left|{\la H_1^{\perp}\ra\over\la D_1\ra}\right|^2
\overline{C}_{TT}{\sin^2\theta_2\over 1+\cos^2\theta_2}$ (in ppm).}
\label{fig4}
\end{wrapfigure}

Meanwhile, the data collected by DELPHI (and other LEP
experiments) give a possibility to measure $H_1^\perp$.  The
point is that despite the fact that the transverse polarization
of quarks in $e^+e^-\to Z^0\to q\bar q$ is very small
($O(m_q/M_Z)$), there is a non-trivial correlation $C^{q\bar
q}_{TT}$ between transverse spins of a quark and an antiquark. In
the Standard Model: $C^{q\bar q}_{TT}=
{(v_q^2-a_q^2)/(v_q^2+a_q^2)}$, which are at $Z^0$ peak:
$C_{TT}^{u,c}\approx -0.74$ and $C_{TT}^{d,s,b}\approx -0.35$.
With the production cross section ratio $\sigma_u/\sigma_d=0.78$
this gives for the average over flavors value $\la
C_{TT}\ra\approx -0.5$.

The transverse spin correlation results in a peculiar azimuthal
angle dependence of produced hadrons, if the T-odd fragmentation
function $H_1^\perp$ does exist~\cite{collins,colpsu}.  A simpler
method has been proposed by Amsterdam group \cite{muldz}. They
predict a specific azimuthal behavior of a hadron in a jet
about the axis in direction of another hadron in the opposite
jet\footnote{The factorized Gaussian form of $p_{h\perp}$
dependence was assumed both for $H_1^{q\perp}$ and $D_1^q$ and
integrated over $|p_{h\perp}|$.}

\begin{eqnarray}
{{\rm d}\sigma\over {\rm
d}\cos\theta_2 {\rm d}\phi_1}\propto (1+\cos^2\theta_2)\cdot \left(1+
{6\over\pi}\left[{H_1^{\perp q}\over D_1^q}\right]^2 C_{TT}^{q\bar
q}{\sin^2\theta_2\over 1+\cos^2\theta_2}\cos(2\phi_1)\right) \, ,
\label{mulders}
\end{eqnarray}
where $\theta_2$ is the polar angle of the electron beam relative
to the second hadron momenta ${\bm p}_2$, and $\phi_1$ is
the azimuthal angle of the first hadron counted off the $({\bm
p}_2,\, {\bm e}^-)$-plane. This asymmetry was probed \cite{todd}
using the DELPHI data collection 91--95.  For the leading charged
particles (mostly pions) in each jet of two-jet events, summed
over $z$ and averaged over quark flavors (assuming
$H_1^{\perp}=\sum_H H_1^{\perp\, q/H}$ is flavor independent),
the most reliable preliminary value of the analyzing power
(obtained from the region $45^\circ<\theta_2<135^\circ$ with
small acceptance corrections) is found to be $\left|{\la
H_1^{\perp}\ra\over\la D_1\ra}\right| =(6.3\pm 2.0)\%$. However,
the larger "optimistic" value
\begin{equation}
\left|{\la H_1^{\perp}\ra\over\la D_1\ra}\right| =(12.5\pm 1.4)\%
\label{apower}
\end{equation}
(obtained from the whole acceptable region
$15^\circ<\theta_2<165^\circ$, see Fig. \ref{fig4}) is not
excluded, with smaller statistical but presumably large
systematic errors. This value, as it will be seen below, better fits
the description of the azimuthal asymmetries in SIDIS.

%%%%%%%%% Section hiQSM %%%%%%%%
\section{Chiral quark-soliton model prediction for
{\boldmath $h_1^a(x)$}}
\label{sect-hiQSM}

\begin{wrapfigure}[29]{TR}{6.0cm}%\begin{figure}[htb]
%\begin{center}
\vspace{-10mm}
\hspace{-4mm}
\includegraphics[width=.37\textwidth]{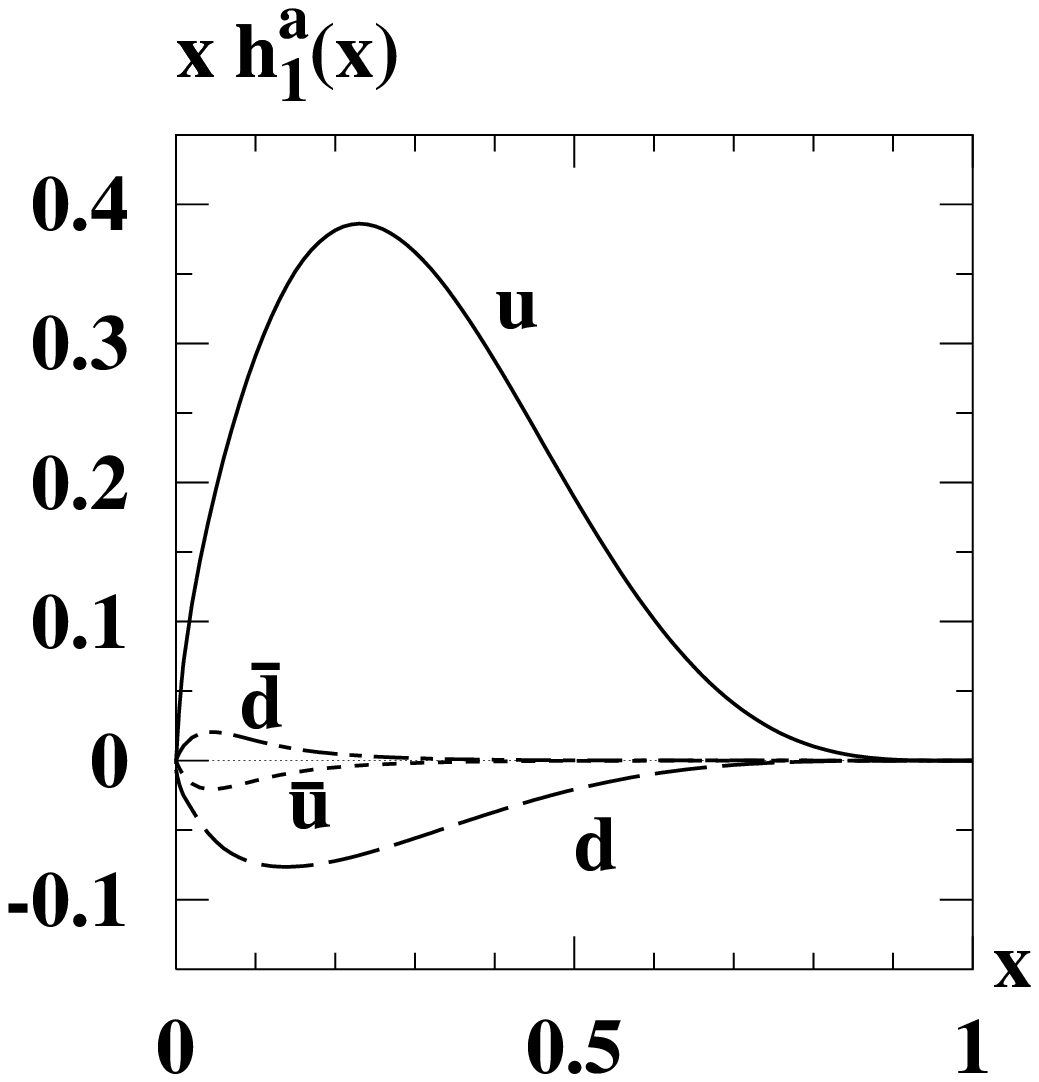}~a)\\[-5mm]
\includegraphics[width=.35\textwidth]{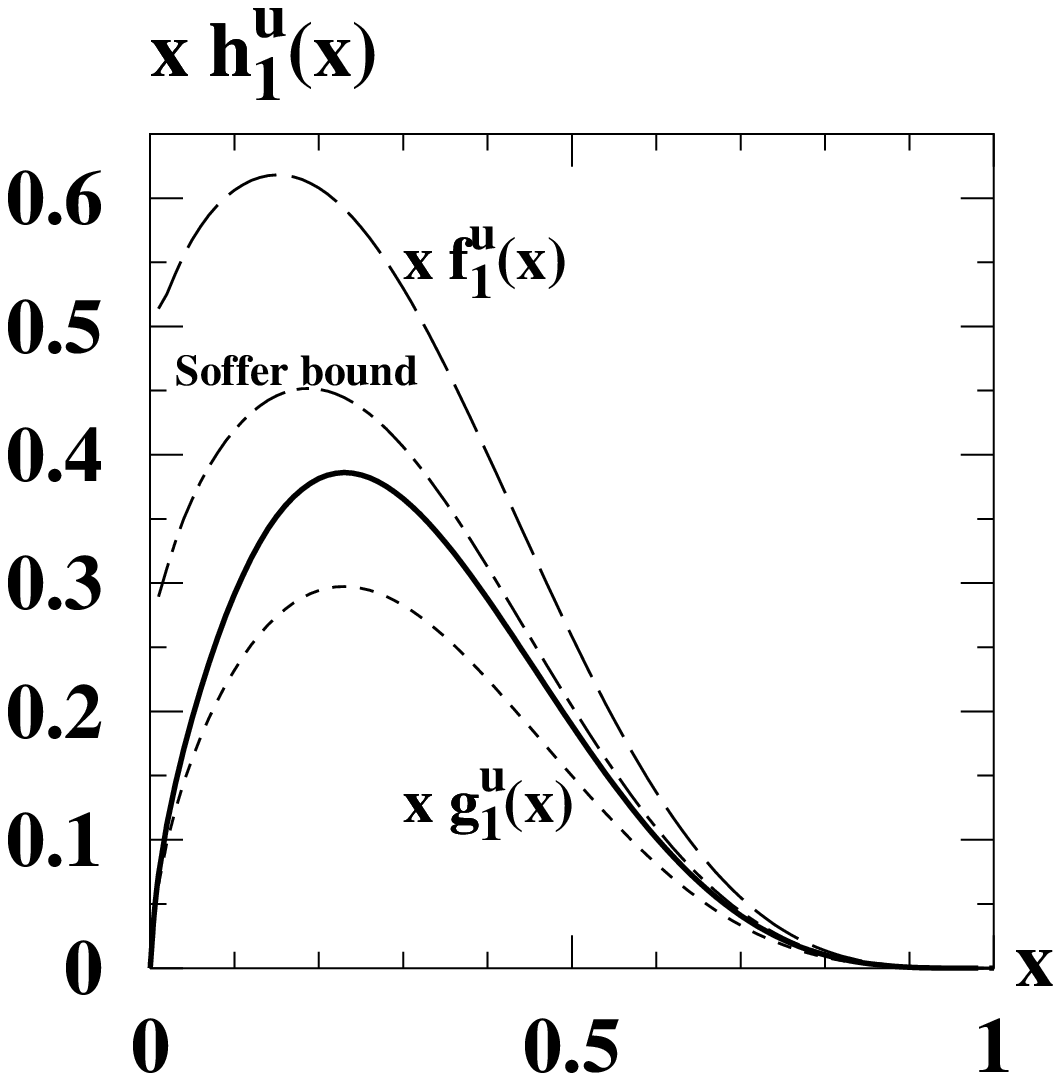}~b)\\[-5mm]
%\end{center}
\caption{{\bf a)}
    The transversity distribution function $h_1^a(x)$ vs.\  $x$
    from the $\chi$QSM
    \cite{Schweitzer:2001sr}.
    {\bf b)}
    Comparison of $h_1^u(x)$ from the $\chi$QSM (solid)
    to $f_1^u(x)$ (dashed) and $g_1^u(x)$ (dotted) and the Soffer bound
    $(f_1^u+g_1^u)(x)/2$ (dashed-dotted line)
    %PS with
    taken or constructed from
    the parameterizations of \cite{Gluck:1994uf}.}
    \label{fig-h1}
\end{wrapfigure}%{figure}

In order to make quantitative estimates for asymmetries we will
use for the transversity distribution function predictions from
the chiral quark-soliton model ($\chi$QSM)
\cite{Schweitzer:2001sr}. This model was derived from the
instanton model of the QCD vacuum \cite{Diakonov:2002fq} and
describes numerous nucleonic properties without adjustable
parameters to within $(10-30)\%$ accuracy \cite{Christov:1995vm}.
The field theoretic nature of the model allows to consistently
compute quark and antiquark distribution functions
\cite{Diakonov:1996sr} which agree with parameterizations
\cite{Gluck:1994uf} to within the same accuracy. This gives us a
certain confidence that the model also describes $h_1^a(x)$ with
a similar accuracy.

In the $\chi$QSM we observe the hierarchy
$h_1^u(x)\gg|h_1^d(x)|\gg|h_1^{\bar u}(x)|$, and an interesting
"maximal sea quark flavour asymmetry" $h_1^{\bar d}(x)\approx
-h_1^{\bar u}(x)>0$. In Fig.~\ref{fig-h1}a we show the $\chi$QSM
prediction for $h_1^a(x)$ from Ref.~\cite{Schweitzer:2001sr}
LO-evolved from the low scale of the model of about
$\mu_0^2=(0.6\,{\rm GeV})^2$ to the scale $Q^2=16\,{\rm GeV}^2$.
In order to gain some more intuition on the predictions we
compare in Fig.~\ref{fig-h1}b the dominating distribution
function $h_1^u(x)$ from the $\chi$QSM to $f_1^u(x)$ and
$g_1^u(x)$ from the parameterizations of
Ref.~\cite{Gluck:1994uf}. It is remarkable that the Soffer
inequality $|h_1^u(x)| \le (f_1^u+g_1^u)(x)/2$ is nearly
saturated -- in particular in the large-$x$ region. (The Soffer
bound in Fig.~\ref{fig-h1}b is constructed from $f_1^u(x)$ and
$g_1^u(x)$ taken at $Q^2=16\,{\rm GeV}^2$ from
\cite{Gluck:1994uf} in comparison with $h_1^u(x)$.)

For the unpolarized distribution function $f_1^a(x)$ we
use the LO parameterization from Ref.~\cite{Gluck:1994uf}.

%%%%%%%%%% Section longpol %%%%%
\section{Comparison with longitudinally polarized data }
\label{sect-longpol}

In the HERMES experiment the $x$- and $z$-dependence of
asymmetries (\ref{asim}) for pions and kaons with $W=\sin\phi$ or
$\sin2\phi$ from longitudinally (relative to the
non-polarized lepton beam) polarized proton and deuterium target
were measured. As it
%PS
is seen from (\ref{pdfpff}) only the second
term of $d\sigma_{0L}$ and the first (Collins) and the third
(Sivers) ones\footnote{\label{chiral-t-odd} Actually, our
approach would imply the vanishing of the Sivers effect. This is
in agreement with the $\chi$QSM. However, this cannot be taken
literally as a prediction for the following reason. The $\chi$QSM
was derived from the instanton vacuum model as the leading order
in terms of the instanton packing fraction $\frac {\rho}{R}\sim
\frac{1}{3}$ ($\rho$ and $R$ are respectively the average size
and separation of instantons in Euclidean space time). In this
order the T-odd PDF's like $f_{1T}^\perp$ and $h_{1}^\perp$
vanish \cite{Pobylitsa:2002fr}. In higher orders the T-odd PDF's
can be well non-zero and all one can conclude at this stage is
that the T-odd PDF's are suppressed with respect to the T-even.
However, considering that $H_1^\perp(z)$ is much smaller than
$D_1(z)$, cf.\ Eq.~(\ref{apower}), it is questionable whether
this suppression could be sufficient such that in physical
cross sections the Collins effect $\propto h_1^a(x)H_1^\perp(z)$
is dominant over the Sivers  effect $\propto
f_{1T}^\perp(x)D_1(z)$. (For an estimation of this suppression
see \cite{Efremov:2003tf,Schweitzer:2003yr}.)} of $d\sigma_{0T}$
with factor (\ref{PT}) do contribute the numerator of
(\ref{asim}) for $W=\sin\phi$ with negative sign ($\phi_S=\pi$)
and only the third one of the $d\sigma_{0L}$ contributes for
$W=\sin2\phi$. This gives for asymmetries

\bea
\label{AUL-sinPhi}
\hspace{-12mm}
A_{0L,P\mbox{\tiny or}D}^{\sin\phi}(x,z,h) &=&
P_{\! L}(x)\;
\frac{\sum_a e_a^2\, x h_L^{a/P\mbox{\tiny or}D}(x)\,H_1^{\perp a}(z)}
{\sum_{a}^h e_{a}^2\, f_1^{a/P\mbox{\tiny or}D}(x)\,D_1^{a}(z)\,}
- P_{\! T}(x)\;
\frac{\sum_a e_a^2\, h_1^{a/P\mbox{\tiny or}D}(x)\,H_1^{\perp a}(z)}
{\sum_{a}^h e_{a}^2\,f_1^{a/P\mbox{\tiny
or}D}(x)\,D_1^{a}(z)\,}\\
\label{AUL-sin2Phi}
\hspace{-12mm}
A_{0L,P\mbox{\tiny or}D}^{\sin2\phi}(x,z,h) &=&
P_{\! 1}(x)\;
\frac{\sum_a e_a^2\, h_{1L}^{\perp(1),a/P\mbox{\tiny or}D}(x)\,H_1^{\perp a}(z)}
{\sum_{a}^h e_{a}^2\, f_1^{a/P\mbox{\tiny
or}D}(x)\,D_1^{a}(z)}\,,
\eea
where $P_{\!L}(x)$ and $P_{\!T}(x)$ are
%PS know
known factors
%PS order of
of order ${\cal O}(M/Q)$ and $P_{\! 1}$
%PS order of
of order ${\cal O}(1)$ depending
on average transverse momenta of partons inside hadron and
hadrons inside parton. The functions $h_L^{a}(x)$ and
$h_{1L}^{\perp(1),a}(x)$ are expressed through transversity
$h_1^{a}(x)$ by the relation (\ref{wwform}).

For $H_1^{\perp a}$ and $D_1^{a}$ a strong suppression of the
unfavoured with respect to the favoured
%PS
fragmentation has been assumed.
From charge conjugation and isospin symmetry one has then
\begin{equation}
\label{H1-favor}
\hspace{-10mm}
H_1^{\perp\rm fav} \equiv
H_1^{\perp u/\pi^+}\!\!=H_1^{\perp\bar d/\pi^+}\!\! =
H_1^{\perp d/\pi^-}\!\! =2H_1^{\perp u/\pi^0}\dots\gg
H_1^{\perp d/\pi^+}\!\! = H_1^{\perp\bar u/\pi^+}\dots
\equiv H_1^{\perp\rm unf}\,.
\end{equation}

\begin{figure}[h!]
\vspace{-3mm}
\begin{center}
\includegraphics[width=4.5cm]{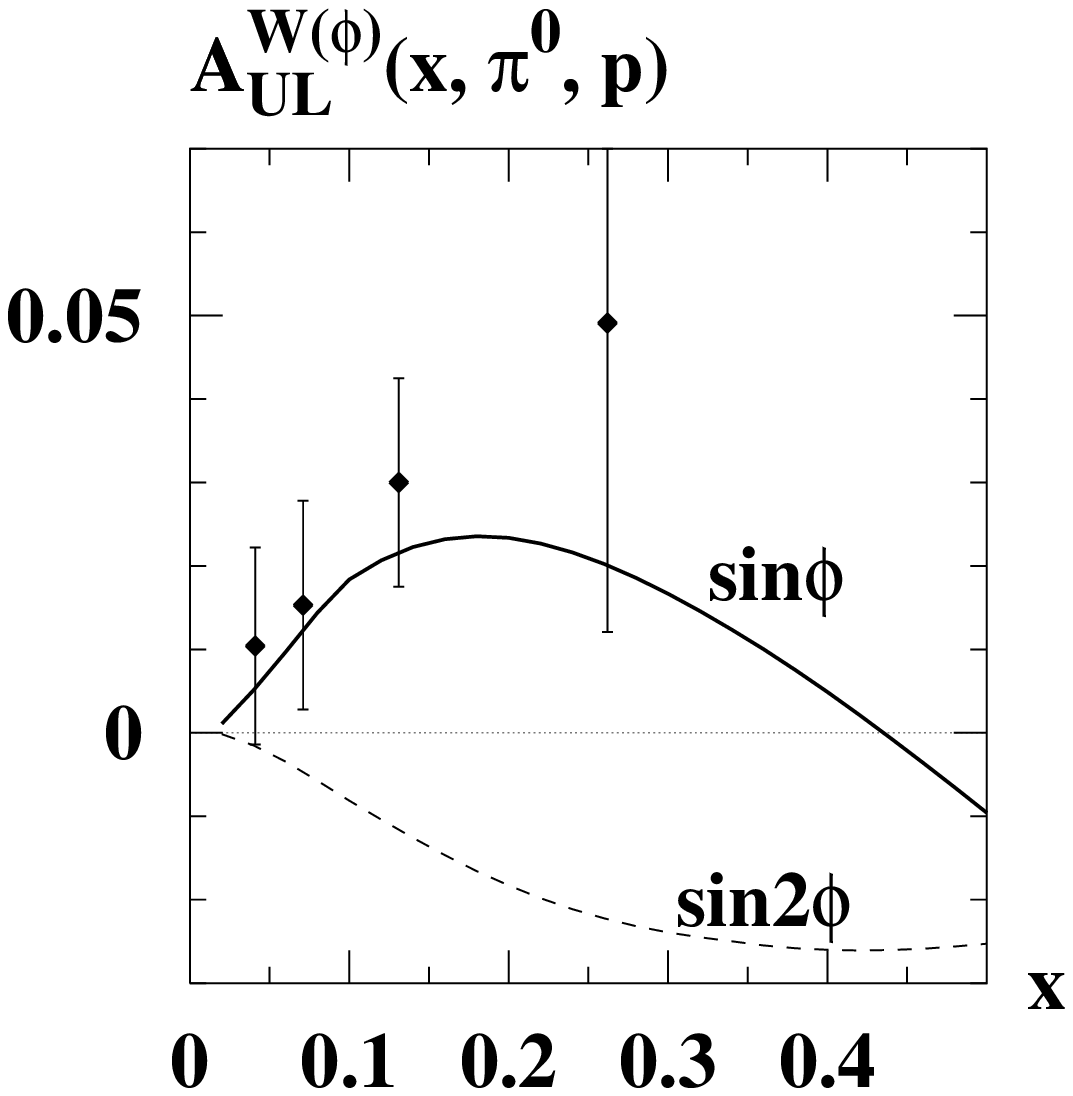}
\includegraphics[width=4.5cm]{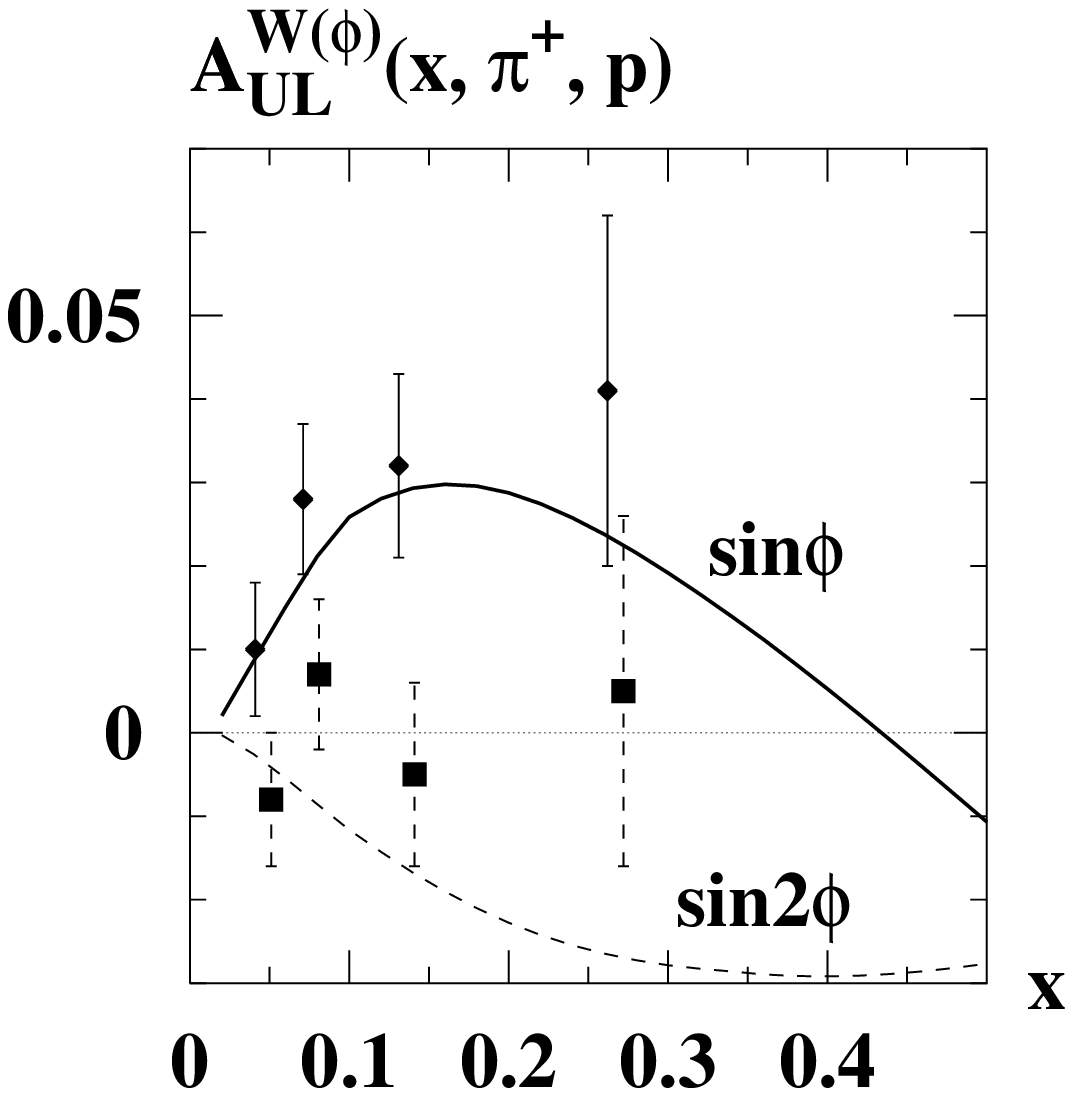}
\includegraphics[width=4.5cm]{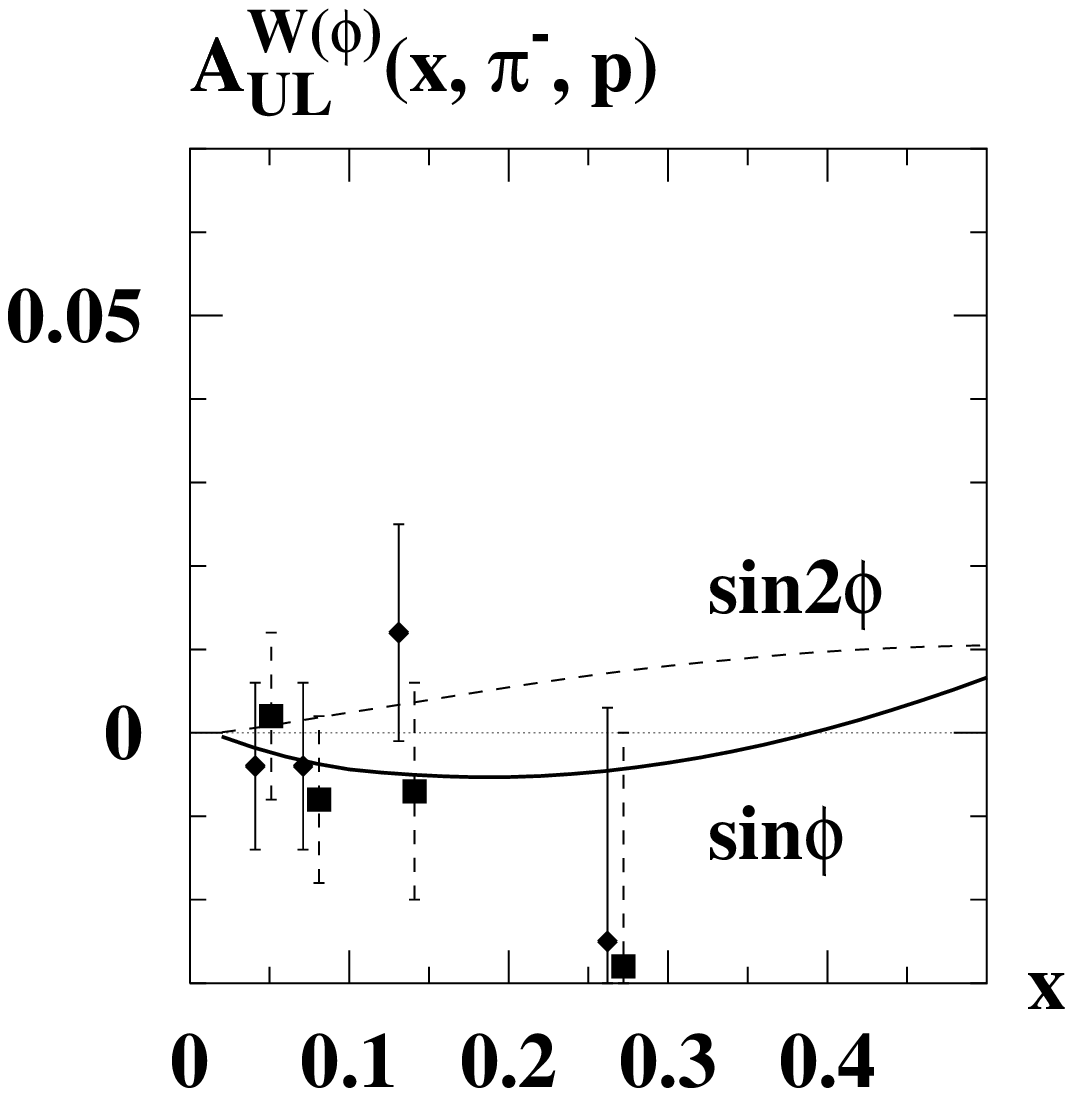}
\end{center}
\vspace{-5mm}
\caption{ Azimuthal asymmetries
$A_{0L}^{W(\phi)}(x,\pi)$ weighted by $W(\phi)=\sin\phi$ (solid
line) and $\sin 2\phi$ (dashed line) for the production of
$\pi^0$, $\pi^+$ and $\pi^-$ as function of $x$. The experimental
data are from Refs. \cite{Airapetian:1999tv,Airapetian:2001eg}.
Rhombus (squares) denote data for $A_{0L}^{\sin\phi}$
($A_{0L}^{\sin2\phi}$). }
\label{AUL-prot}
\end{figure}

So, using the DELPHI result\footnote{ We assume a  weak scale
dependence of the analyzing power (\ref{apower}).}
Eq.(\ref{apower}), $\chi$QSM for $h_1^a(x)$, the relation
(\ref{wwform}) for the $h_L$ and the parameterization from Ref.
\cite{Gluck:1994uf} for $f_1^a(x)$, both LO-evolved to the
average scale $Q_{\rm av}^2=4\,{\rm GeV}^2$ characteristic for
HERMES we obtain for the proton target Fig. \ref{AUL-prot}
%%%--- FIGURE 2:  Analyzing power as function of z -------
\begin{figure}[t]
\begin{minipage}{70mm}
\vspace{-2mm}
\includegraphics[width=1\textwidth]{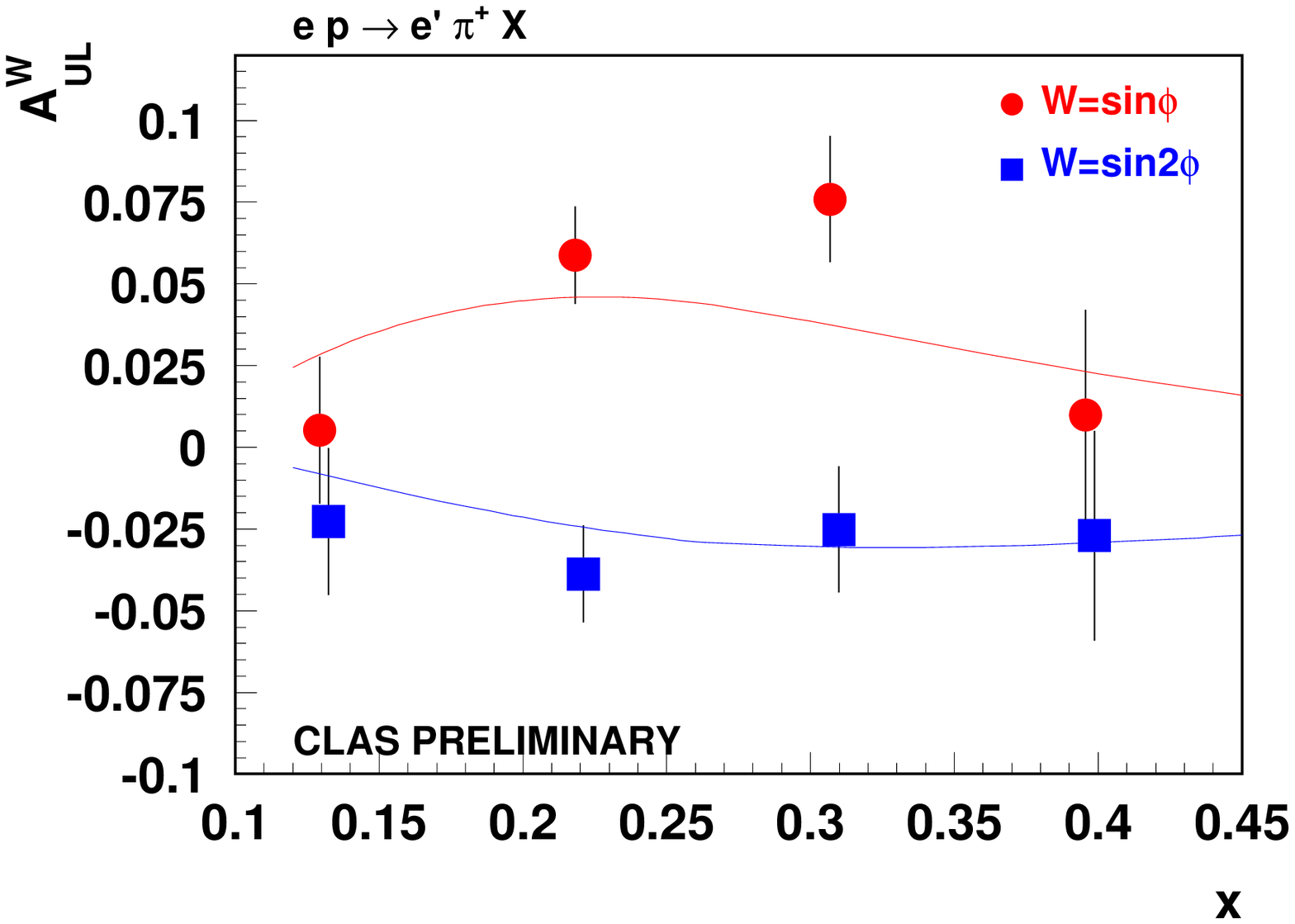}\\[-8mm]
\caption{Our predictions (solid curves) for azimuthal asymmetries
$A_{0L,D}^{\sin2\phi}$ and $A_{0L,D}^{\sin\phi}$ vs. $x$ in
comparison with CLAS data \cite{Avakian:2002qp}.}
\label{AUL-sin2phi}
\end{minipage}
\hfill
\begin{minipage}{70mm}
\vspace{-4mm}
\includegraphics[width=.9\textwidth,height=.24\textheight]
{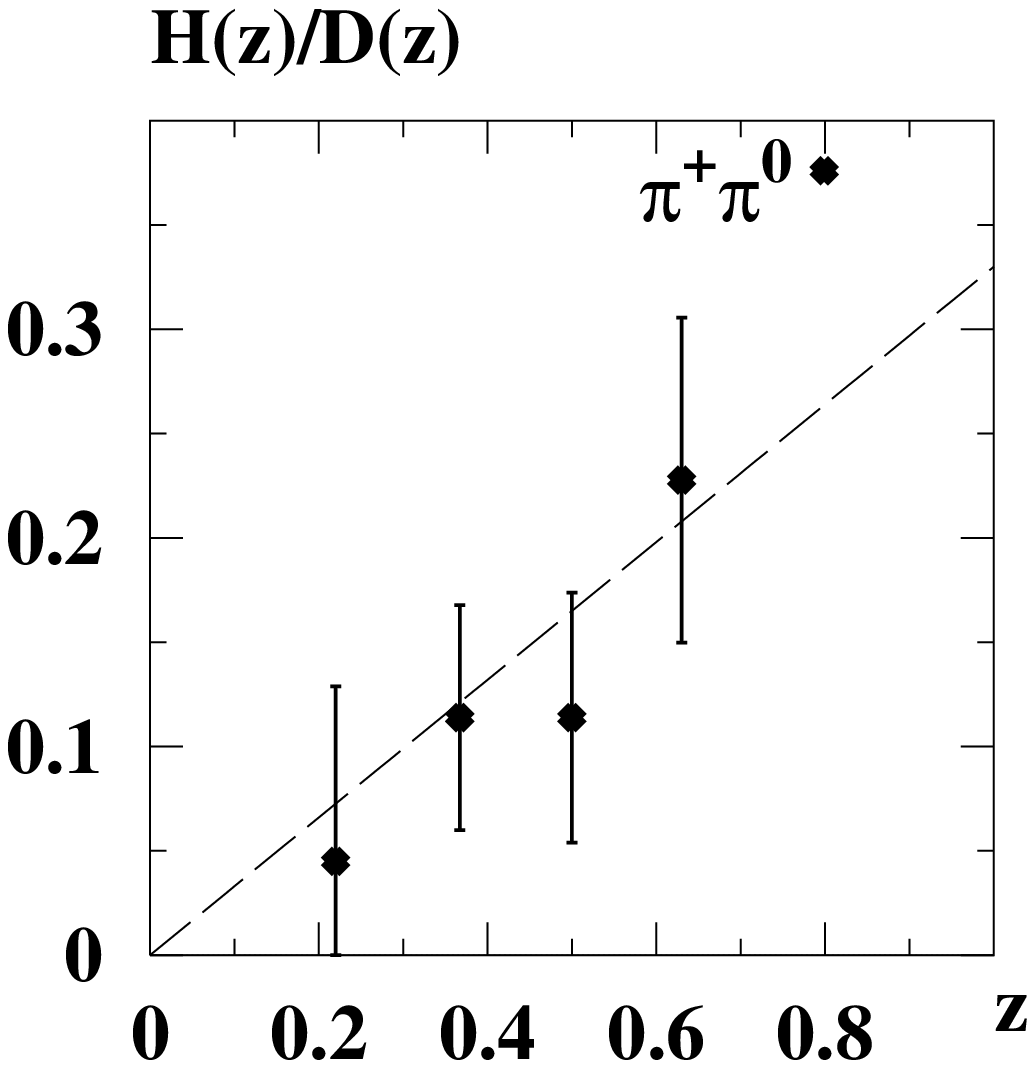}\\[-5mm]
\caption{$H_1^\perp/D_1$
vs. $z$, as extracted from HERMES data
\cite{Airapetian:1999tv,Airapetian:2001eg}
for $\pi^+$ and $\pi^0$ production combined.}
\label{fig-H(z)}
\end{minipage}
\end{figure}

\begin{wrapfigure}[27]{RT}{5.0cm}
\vspace{-7mm}
\includegraphics[width=.35\textwidth,height=.51\textheight]
{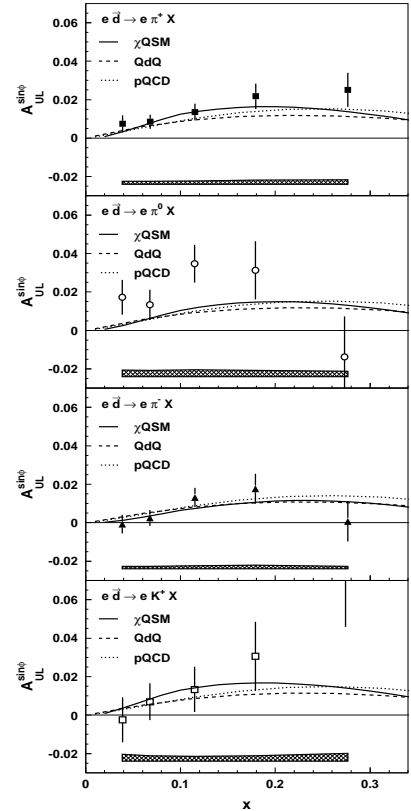}\\[-7mm]
\caption{
Our predictions (solid curves) for
$A_{0L,D}^{\sin\phi}(x,h)$
vs. $x$ from a longitudinally polarized deuteron target in
comparison with HERMES data \cite{Airapetian:2002mf}.}
\label{AUL-deut-sinphi}
\end{wrapfigure}
%%%--- END FIGURE 2. --------------------

Fig. \ref{AUL-sin2phi} shows also our predictions for
$A_{0L,P}^{\sin\phi}(x,\pi)$ and
$A_{0L,P}^{\sin2\phi}(x,\pi)$ in comparison with CLAS data
\cite{Avakian:2002qp}
%PS at 5.7 GeV.
for a 5.7 GeV beam.

We conclude that the azimuthal asymmetries obtained with the
$\chi$QSM prediction for $h_1^a(x)$ \cite{Schweitzer:2001sr}
combined with the ``optimistic'' DELPHI result (\ref{apower}) for
the analyzing power are consistent with experiment {\sl with no
fit parameter}.

It is interesting to note that the negative sign of the
transversal contribution in (\ref{AUL-sinPhi}) leads to a change
of sign of the $A^{\sin\phi}_{0L}$ asymmetries for $x>0.4$. This
is due to a harder behaviour $h_1(x)$ with respect to $h_L(x)$
followed from Eq. \ref{wwform} . It should be noted that the
prediction of $A_{0L}^{\sin\phi}(x,\pi)=0$ at $x\simeq(0.4-0.5)$
is sensitive to the approximation of favoured flavour
fragmentation (\ref{H1-favor}). In principle one could conclude
from data, how well this approximation works. However, the upper
$x$-cut is $x<0.4$ in the HERMES experiment
\cite{Airapetian:1999tv,Airapetian:2001eg}.

Now let us reverse the logic, and using z-dependence of the
HERMES results for the $\pi^0$ and $\pi^+$ azimuthal asymmetries
try to estimate $H_1^\perp(z)/D_1(z)$. For that we use the
$\chi$QSM prediction for $h_1^a(x)$ that will introduce a model
dependence of order (10 - 20)\%. The combined result is shown in
Fig. \ref{fig-H(z)}.  The data can be described by a
linear fit\\
\begin{minipage}{95mm}
\be
H_1^\perp(z)=(0.33 \pm  0.06)zD_1(z)
\label{apower-vs-z}
\ee
\end{minipage}\\[3mm]
with average ${\la H_1^\perp\ra}/{\la D_1\ra}=(13.8\pm 2.8)\%$
which is in good agreement with DELPHI result Eq.(\ref{apower}).
The errors are the statistical errors of the HERMES data. It is
interesting to note that numerically the behavior
(\ref{apower-vs-z}) is close to those calculated from chirally
invariant Manohar-Georgi model \cite{Bacchetta:2002es},
$H_1^\perp(z)\approx 0.63z^2D_1(z)$.

Believing that such acceptable description  of the proton data is
not occasional we made the predictions \cite{Efremov:2001ia}
for $A_{0L}$ asymmetries for pions and kaons at
longitudinally polarized deuteron target which
%PS  were measuring at those time
were being measured at that time by the HERMES collaboration.

The main question
%PS  was however of
was, however, how large is the analyzing power
for kaons? We know that the unpolarized kaon fragmentation
function $D_1^K(z)$ is roughly five times smaller than the
unpolarized pion one. Is also $H_1^{\perp\,K}(z)$ five times
smaller than $H_1^{\perp\,\pi}(z)$? {The reason is that in chiral
limit $D_1^\pi=D_1^K$ and $H_1^{\perp\pi}=H_1^{\perp K}$. The
naive expectation is that the 'way off chiral limit to real
world' proceeds analogously as for spin-dependent quantities,
$H_1^\perp$, as for spin-independent one, $D_1$.}
If we assume this, i.e. if%\\
%\begin{minipage}{95mm}
\be
\frac{\la H_1^{\perp\,  K}\ra}{\la D_1^K  \ra} \simeq
\frac{\la H_1^{\perp\,\pi}\ra}{\la D_1^\pi\ra}
\label{apower-kaon}
\ee
%\end{minipage}\\[3mm]
holds, we obtain -- with the central value of $\la
H_1^\perp\ra/\la D_1\ra$ in Eq.(\ref{apower}) -- azimuthal
asymmetries for $K^+$ and $K^0$ as large as for pions. The
results of our predictions \cite{Efremov:2001ia} (solid curves)
in comparison with the published HERMES data
\cite{Airapetian:2002mf} are presented at Fig.
\ref{AUL-deut-sinphi}. Again no fit parameters
%PS  was
were used in distinction with other models at Fig.
\ref{AUL-deut-sinphi}. The asymmetries for $\bar K^0$ and $K^-$
are close to zero in our approach.

\begin{wrapfigure}[16]{RH}{5.0cm}
%\vspace{-3mm}
%\includegraphics[width=.35\textwidth,height=.2\textheight]{figure6.eps}\\[-5mm]
%\caption{Our predictions (solid curves) for azimuthal asymmetries
%$A_{0L,D}^{\sin2\phi}$ vs. $x$ in comparison with HERMES data
%\cite{Airapetian:2002mf}.}
%\label{AUL-deut-sin2phi}
%\end{wrapfigure}
%\begin{wrapfigure}[]{RH!}{5.0cm}
\vspace{-10mm}
\includegraphics[width=.35\textwidth]
{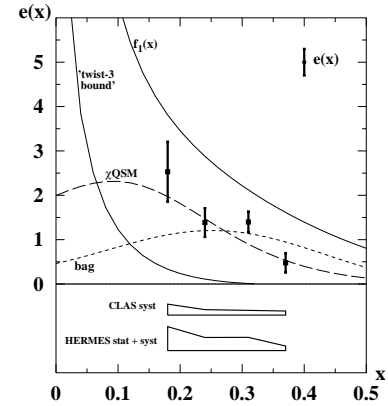}\\[-5mm]
\caption{
The flavour combination
$e(x)\!=\!(e^u\!+\!\frac{1}{4}e^{\bar d})(x)$, with error bars
due to statistical error of CLAS data, vs. $x$ at $\la Q^2\ra
\!=\! 1.5\,{\rm GeV}^2$. For comparison the twist-3 bound,
$f_1^u(x)$, bag and $\chi$QSM models predictions are shown. }
\label{fig-eofx}
\end{wrapfigure}
%====== SECTION 5: CLAS AND EXTRACTION OF e(x) =====================
\section{Extraction of {$\mathbf e(x)$} from $\mathbf{A_{L0}}$
}

Very recently the $\sin\phi$ asymmetry of $\pi^+$ produced by
scattering of polarized electrons off unpolarized protons was
published by CLAS collaboration \cite{Avakian:2003pk} and
preliminary data were reported by HERMES collaboration
\cite{Avetisyan:2004uz}. This
asymmetry is interesting since it allows to access the unknown
twist-3 structure functions $e^a(x)$ (see $d\sigma_{L0}$ in
(\ref{pdfpff})) that are connected with the nucleon $\sigma$-term:\\
\begin{minipage}{93mm}
\be
\label{sigma}
    \int_0^1\di x\sum_ae^a(x)=\frac{2\sigma}{m_u+m_d}
    \approx 10 \; .
\ee
\end{minipage}\\[3mm]

The asymmetry is given by\footnote{The term $\propto \sum_a
e_a^2h_1^{\perp a}(x)\la E^{a/\pi}\ra$ in numerator was
disregarded since $h_1^{\perp a}$ are suppressed in $\chi$QSM
(see footnote \ref{chiral-t-odd}.)}

\begin{minipage}{89mm}
\be
\hspace{-3mm}
    A_{L0}^{\sin\phi}(x)\propto \frac{M}{Q} \;
    %\frac{4y\sqrt{1-y}}{1+(1-y)^2}
    \frac{\sum_a e_a^2e^a(x)\la H^{\perp a/\pi}_1\ra}
    {\sum_a e_a^2 f^a_1(x)\la D_1^{a/\pi}\ra}  \;.
\label{ALU}
\ee
\end{minipage}\\[3mm]

Disregarding unfavored fragmentation and using the Collins
analysing power extracted from HERMES in (\ref{apower-vs-z}) that
yields for $z$-cuts of CLAS ${\la H^{\perp\pi}_1\ra}/{\la
D_1^\pi\ra}=0.20 \pm 0.04$, we can extract \cite{Efremov:2002ut}
$e^u(x)+\frac14e^{\bar d}(x)$. The result is presented in Fig.
\ref{fig-eofx}. For comparison the twist-3 lower
bound\footnote{\label{footnote-Soffer}Let us stress that strictly
speaking this inequality could be justified only if the ``twist-2
Soffer inequality'' $2|h_1^a(x)|\le(f_1^a+g_1^a)(x)$ of
Ref.~\cite{Soffer:1995ww} were saturated \cite{Efremov:2002qh}.
In the following we will refer to this relation as ``twist-3
lower bound'' keeping in mind that it does not need to hold in
general. In Ref.~\cite{Lu:1996ae} a bound based on the positivity
of the hadronic tensor and the Callan-Gross relation (and
formulated in terms of structure functions) was discussed.},
$e^a(x)\ge 2|g_T^a(x)|-h_L^a(x)$ Ref.~\cite{Soffer:1995ww}, and
the unpolarized distribution function $f_1^u(x)$ are plotted. The
prediction of $\chi$QSM \cite{Schweitzer:2003uy} and bag model
\cite{Signal:1997ct} are shown also. One can guess that the large
number in the sum rule Eq.(\ref{sigma}) might be due to, by all
means, a $\delta$-function at $x=0$. (For review and references
see \cite{Efremov:2002qh}.)

%%%%%%%%%%%%%%% Section transpol %%%%%%%%%%%%

\section{%Collins effect for
SIDIS with transverse target polarization} \label{sect-transpol}

In the HERMES and COMPASS experiments the cross sections
$\sigma_N^{\uparrow\downarrow}$ for the process
$lN^{\uparrow\downarrow}\rightarrow l'h X$ is measured, where
$N^{\uparrow\downarrow}$ denotes the transversely relative to
the beam polarized target (see Fig.~\ref{kinsidis}).

 The component of the target polarization vector which is
{\sl transverse relative to the hard photon} is characterized
by the angle $\Theta_S$, see Fig.~\ref{kinsidis}, given by
\be
\label{spin-projection}
\hspace{-8mm}
\sin\Theta_S
=\cos\theta_\gamma\sqrt{1+{\rm tan}^2\theta_\gamma\sin^2\phi_{S'}}
\approx \cos\theta_\gamma \approx 1\;,
\ee
where $\phi^\prime_S$ is the azimuthal angle of the target
polarization direction about the lepton beam direction relative
to the scattering plane. As it
%PS
is seen from (\ref{pdfpff}) only
the first (Collins) term of $d\sigma_{0T}$ with factor
(\ref{spin-projection})
%PS  do contribute
does contribute to
the numerator of (\ref{asim})
for $W=\sin(\phi+\phi_S)$. This gives for Collins asymmetry
\be
A_{0T}^{\sin(\phi+\phi_s)}(x,z,h) = B_T(x)\;
\frac{\sum_a e_a^2\,x\, h_1^a(x)\,H_1^{\perp a}(z)}
{\sum_b e_b^2\,x\, f_1^b(x)\,D_1^b(z)\,} \;,
\label{AUT-without-kT-fin}
\ee
where $B_{\!T}(x)$ are known factors order of ${\cal O}(1)$
depending on average transverse momenta.

\begin{figure}[t]
%\vspace{-0.3cm}
\begin{center}
\includegraphics[width=.30\textwidth]
{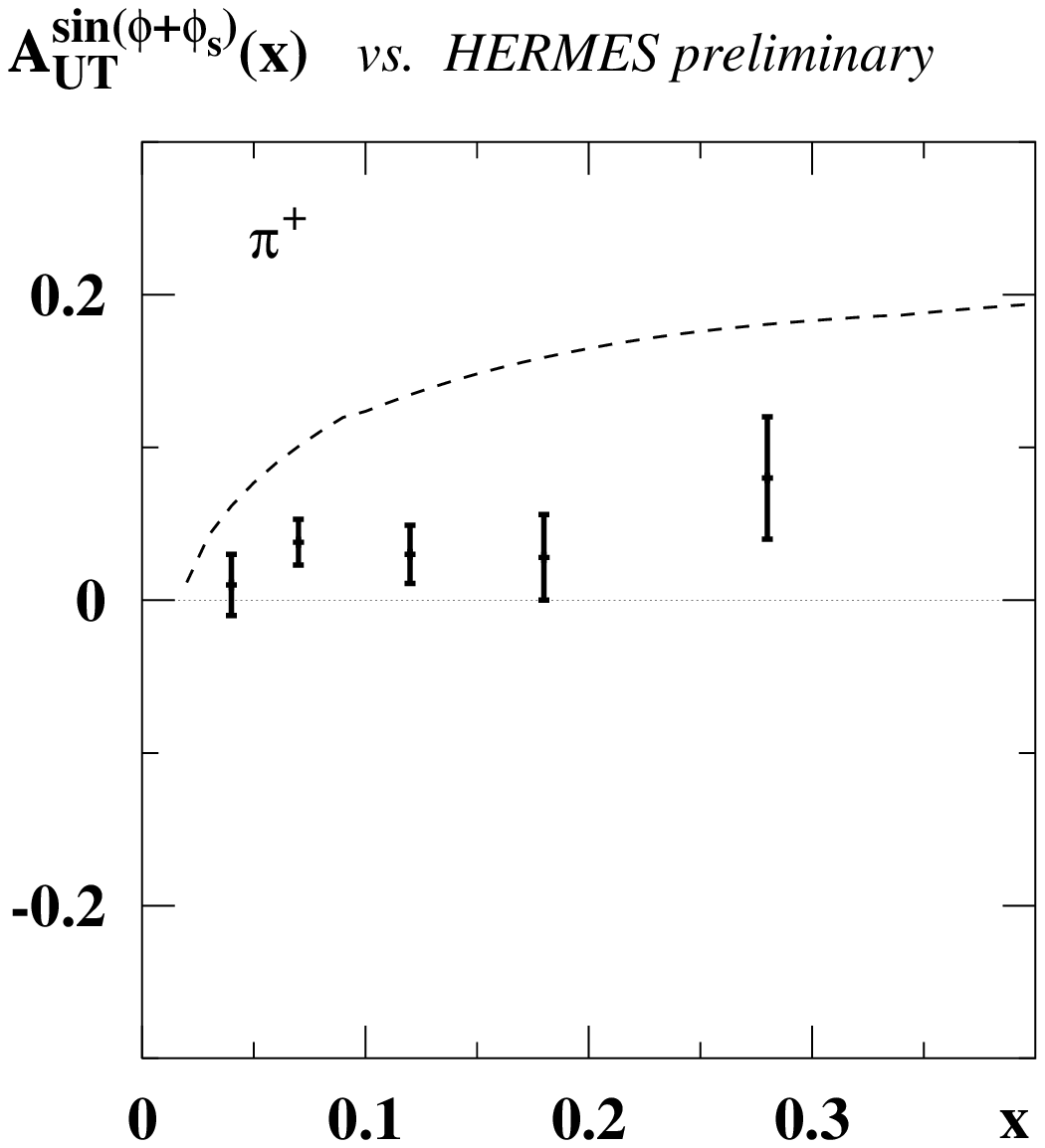}
\includegraphics[width=.30\textwidth]
{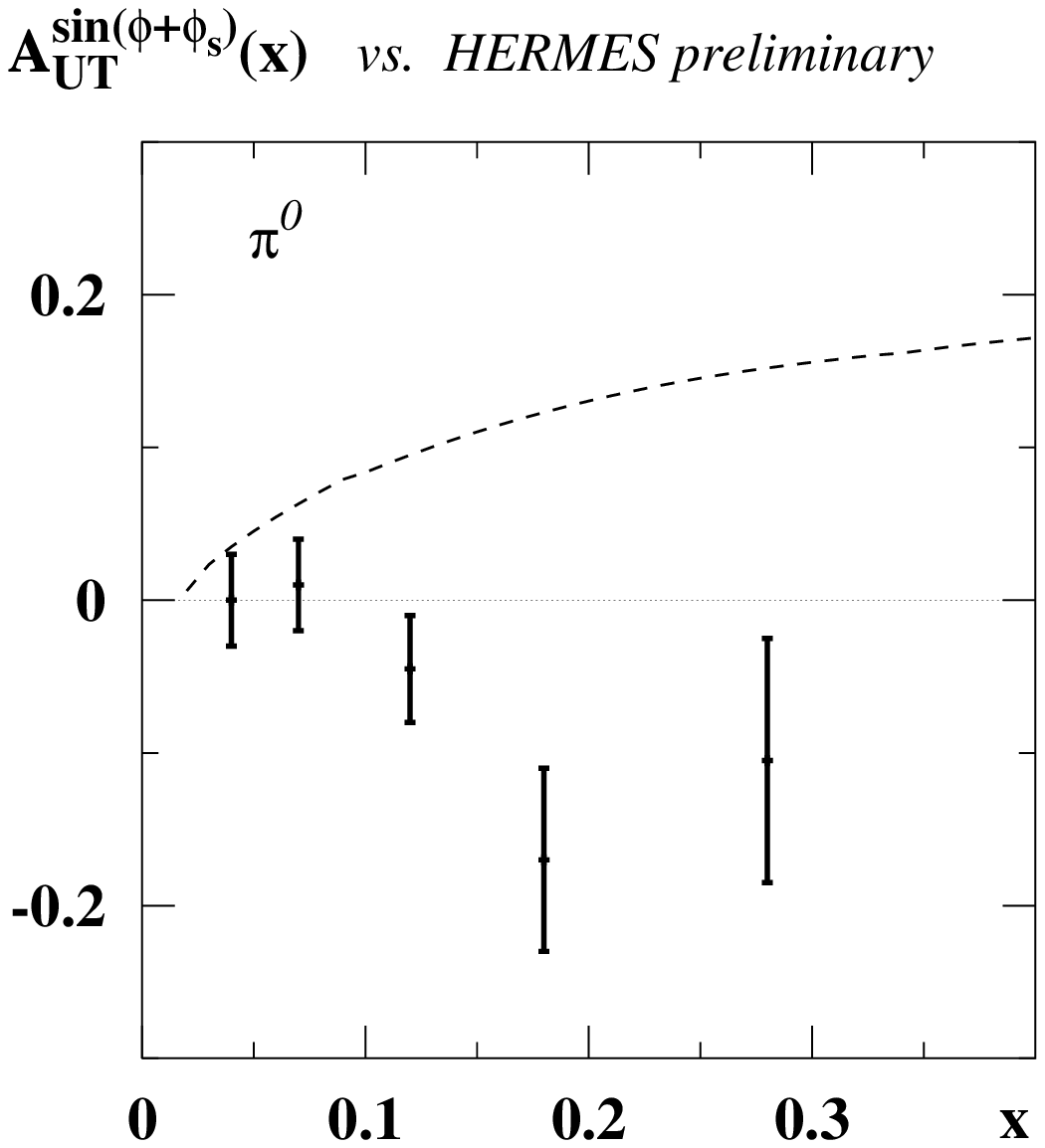}
\includegraphics[width=.30\textwidth]
{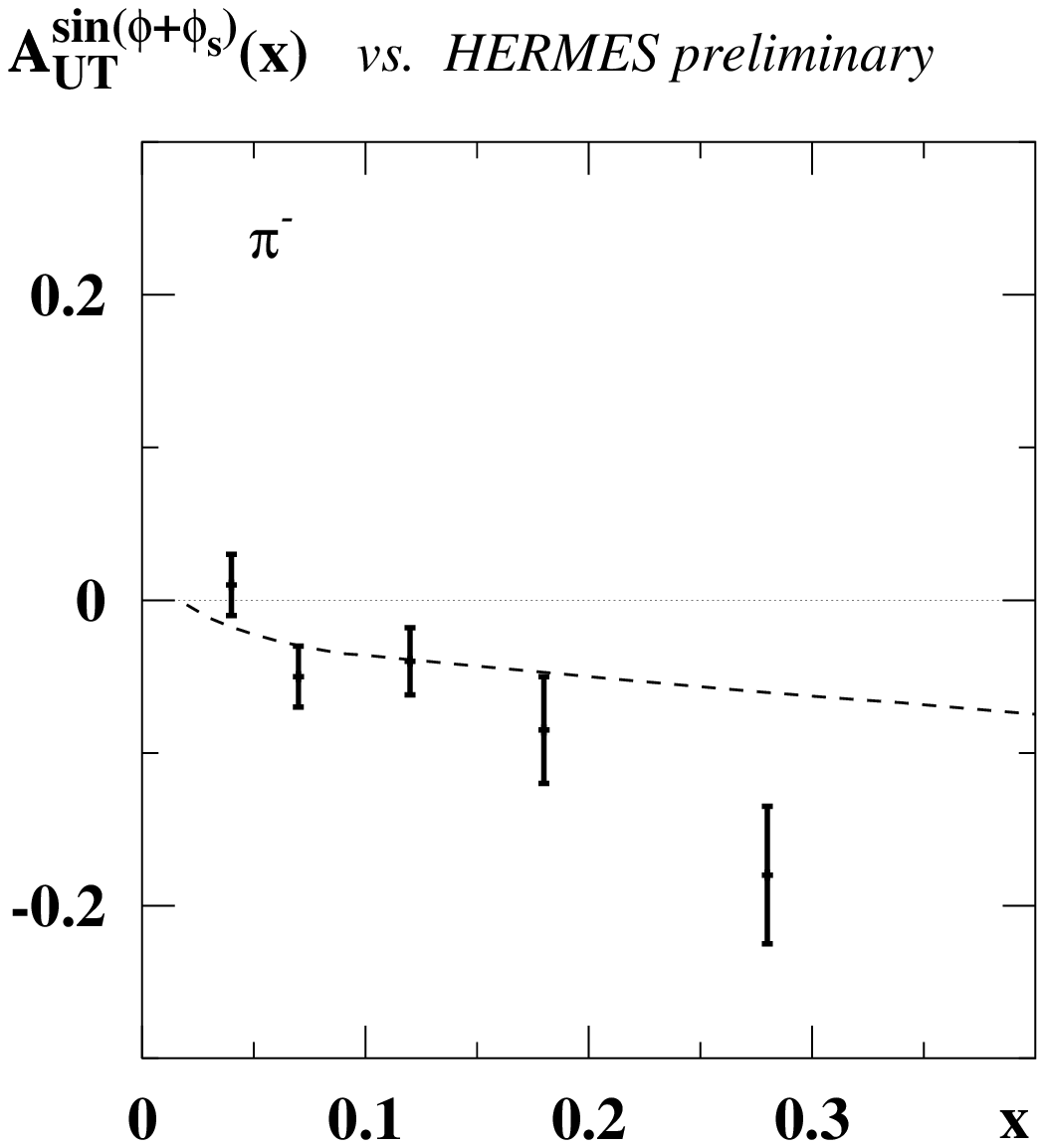}
\end{center}
\vspace{-0.3cm}
\caption[*]{
The Collins effect transverse target SSA $A_{0T}^{\sin(\phi+\phi_s)}$
in the production of $\pi^+$, $\pi^0$ and $\pi^-$ from a proton target.
{\sl Preliminary} data are from \cite{AUT-HERMES},
theoretical curves from \cite{Efremov:2003eq}.}
\label{AUT-col-HERMES}
\end{figure}

Based on our understanding of the longitudinally polarized target
asymmetry, predictions for the Collins effect for transversally
polarized target asymmetry $A_{0T}^{\sin(\phi+\phi_s)}$
%PS was
were made\footnote{Asymmetries of similar
%PS values
magnitude as for HERMES are
also predicted \cite{Efremov:2003eq} for
%PS
the running COMPASS
experiment with transversally and longitudinally polarized
targets.} \cite{Efremov:2003eq} (see Fig. \ref{AUT-col-HERMES}).
Of course, since the theoretical description of the power
suppressed (``twist-3'')  longitudinal asymmetry is involved and
we made simplifications, which are difficult to control, one
cannot expect that we accurately predict the overall magnitude of
the effect. However, one could have a certain confidence that at
least {\sl the sign and the shape} of
$A_{0T}^{\sin(\phi+\phi_s)}(x)$ is described satisfyingly since
it is dictated by the model prediction for $h_1^a(x)$
\cite{Schweitzer:2001sr} and the approximation of favoured
flavour fragmentation only. As can be seen in Fig.
\ref{AUT-col-HERMES} our results \cite{Efremov:2003eq} do not
even describe the sign for $\pi^0$ of the preliminary HERMES data
\cite{AUT-HERMES}! Why not?

Apparently some assumption(s) made must be incorrect. The
first suspicion is favoured fragmentation approximation (\ref{H1-favor}).
First of all, pay attention
%PS for
to the negative and large
$A_{0T}^{\sin(\phi+\phi_s)}(\pi^0)$. With the unfavored
fragmentation taken into account one
%PS have
has from charge conjugation
and isospin invariance (\ref{H1-favor})
\begin{equation}
\label{pi0-frag}
%\hspace{-10mm}
H_1^{\perp a/\pi^0}=
\frac{1}{2}(H_1^{\perp a/\pi^+}+H_1^{\perp a/\pi^-})
=\frac{1}{2}(H_1^{\perp\rm fav}+ H_1^{\perp\rm unf})\,.
\end{equation}
Then
\be
\label{AUT-unf-pi0}
\underbrace{A_{0T}^{\sin(\phi+\phi_s)}(\pi^0)}_{<0\;\rm
in\;experiment} \propto \underbrace{{\textstyle\sum_a}e_a^2
h_1^a(x)}_{>0\;\rm in\; models} \; \la H_1^{\perp\rm
fav}+H_1^{\perp\rm unf}\ra \;\;\; \Longrightarrow \;\;\; \la
H_1^{\perp\rm fav}+H_1^{\perp\rm unf}\ra <0 \;.
\ee

In order to explain
%PS
the asymmetry for charged pions the option
$H_1^{\perp \rm fav}<0$ can be ruled out, unless
$(4h_1^u+h_1^{\bar d})<(h_1^d+4h_1^{\bar u})$ which would
contradict any
%PS models.
model. Then, with option $H_1^{\perp \rm fav}>0$,
we can draw two interesting conclusions from the observation in
Eq.~(\ref{AUT-unf-pi0}). Firstly, $H_1^{\perp \rm unf}$ should
have opposite sign with respect to $H_1^{\perp \rm fav}$. This
could have a natural explanation in string models, in particular
for the HERMES kinematics \cite{AUT-HERMES} with low particle
multiplicity jets. Secondly, the absolute value of $H_1^{\perp
\rm unf}$ has to be larger than the absolute value of $H_1^{\perp
\rm fav}$ which, if confirmed, will be more difficult to
understand.

Concerning the large value of $A_{0T}^{\sin(\phi+\phi_s)}(\pi^0)$
notice that it is approximately of the same order as
$A_{0T}^{\sin(\phi+\phi_s)}(\pi^-)$. Meanwhile from the
factorization of $x$ and $z$ dependence of polarized and
unpolarized SIDIS cross sections and from relation
(\ref{pi0-frag}) one can write for {\sl any} spin asymmetry \be
\hspace{-10mm}
A(\pi^0)=\frac{\sigma(\pi^+)}{\sigma(\pi^+)+\sigma(\pi^-)}\,A(\pi^+)
+\frac{\sigma(\pi^-)}{\sigma(\pi^+)+\sigma(\pi^-)}\,A(\pi^-)
=A(\pi^-)+a[A(\pi^+)-A(\pi^-)]\,, \ee where $\sigma(\pi)$ is
unpolarized SIDIS cross section. If $A(\pi^0)\approx A(\pi^-)$
then $a=\frac{\sigma(\pi^+)}{\sigma(\pi^+)+\sigma(\pi^-)}\approx
0$ that is nonsense! This leads to conclusion that the
factorization of $x$ and $z$ dependence for the transversally
polarized SIDIS cross sections is under suspicion and should be
carefully checked. Regrettably the statistical errors are rather
large, especially for $A_{0T}(\pi^0)$. Probably due to this
reason the final publication \cite{unknown:2004tw} do not contain
$A_{0T}(\pi^0)$. But even without $\pi^0$ the question stays
why $A_{0T}(\pi^+)$ is so small and $|A_{0T}(\pi^-)|$ is so
large since unfavored $\la H_1^{\perp\rm unf}\ra$ gives opposite
sign contributions to $\pi^+$ and $\pi^-$ asymmetries the latter
about order of magnitude larger.

The present situation seems paradoxical. We have a reasonable
understanding of $A_{0L}$ asymmetries, but know that it possibly
%PS are
is based on an incomplete theoretical description of the process
-- with the Sivers effect and other contributions omitted. We
probably have a more reliable description of the Collins $A_{0T}$
asymmetry, but cannot understand the {\sl preliminary} data.

However, one should keep in mind the preliminary stage of the
data \cite{AUT-HERMES}, which does not allow yet to draw more
definite conclusions. Further data from HERMES as well as
COMPASS, CLAS, HALL-A and HALL-B experiments will contribute
considerably to resolve the present puzzles and pave the way
towards a qualitative understanding of the numerous new
distribution and fragmentation functions.

\section{Conclusion}

Processes with polarized particles have always been among the
most difficult and complicated themes both for experimentalists
and theorists.

First, working with polarized targets, experimentalists have to
"battle with" thermal chaos which tends to break the polarized
order. For this one needs liquid helium temperatures. More
difficulties, like depolarizing resonances, are encountered in
accelerating polarized particles and in controlling a polarized
beam. Second, spin effects are very perfidious: as a rule, they
are strongest in kinematical regions where the process itself is
the least probable.

As for the theory, I can hardly recall a case when its first
prediction was correct! As a rule, it was wrong and forced
theorists to think more fundamentally to repair the theory. This
resulted in a deeper understanding of particle interaction
mechanics. Nevertheless many puzzles such as "Why are hyperons
produced so strongly polarized?" or "What is the structure of the
nucleon spin?" stay yet unsolved during decades.
%PS Do the single spin asymmetries will became
Will single spin asymmetries become one more such problem?
The future will show.

\begin{acknowledgement}
The author is partially supported by grants
RFBR 03-02-16816 and DFG-RFBR 03-02-04022.
\end{acknowledgement}

\end{document}